\documentclass[11pt]{article}
\usepackage{moriond,epsfig}
\usepackage{subfigure}
\usepackage{amsmath}

\def\be{\begin{equation}}
\def\ee{\end{equation}}
\def\bea{\begin{eqnarray}}
\def\eea{\end{eqnarray}}

\begin{document}
\vspace*{4cm}
\title{TOP-QUARK MASS MEASUREMENTS AT THE LHC}

\author{\textsc{S. Blyweert}, on behalf of the ATLAS and CMS Collaborations}

\address{Vrije Universiteit Brussel - Interuniversity Institute for High Energies\\
Pleinlaan 2, 1050 Brussel, Belgium}

\maketitle\abstracts{
The top-quark mass $m_{\rm t}$ is one of the fundamental parameters of the standard model (SM) of particle physics. The CDF and D0 collaborations already performed very precise measurements of $m_{\rm t}$. Since the ATLAS and CMS collaborations already have a very large sample of top quark pairs available for analysis, they are producing results with increasing precision. An overview of the most recent measurements of $m_{\rm t}$ by ATLAS and CMS is given, using up to 4.7fb$^{-1}$ of data in different $\rm t \bar{t}$ decay channels. The measurement of the mass difference between top and antitop quarks is also shown. Finally, the combination of the individually measured $m_{\rm t}$ values is discussed, together with an outlook for future $m_{\rm t}$ measurements.
}

\section{Introduction}

The top quark, which was discovered in 1995 at the Tevatron, is the heaviest currently known fundamental particle. Its mass is an important parameter of the standard model, since it is an important input for the global Electro-Weak fits. These fits can be used to constrain the mass of the SM Brout-Englert-Higgs boson, and are also a consistency check of the standard model. $m_{\rm t}$ has been measured already by the CDF and D0 collaborations with great precision, resulting in $m_{\rm t} = 173.2 \pm 0.9$ GeV as the current world average~\cite{TevatronCombi,TevTopMassTalk}. A precise measurement of $m_{\rm t}$ by the ATLAS and CMS collaborations at the Large Hadron Collider (LHC) would provide an independent cross-check of this value and would help to further reduce the total uncertainty on the world average of $m_{\rm t}$.

In proton-proton collisions at $\sqrt{s} = 7$~TeV top quarks are dominantly produced in pairs. In 2011 the LHC delivered more than 5~fb$^{-1}$ to both the ATLAS and CMS experiments, corresponding to about $8 \cdot 10^5$ ${\rm t\bar{t}}$ pairs per experiment. This gave both ATLAS and CMS the opportunity to perform precise measurements of $m_{\rm t}$ in the different decay channels. Since top quarks decay most of the time into a b quark and a W boson, ${\rm t}\bar{\rm t}$ events can be categorized, according to the decay of the W bosons, in dileptonic (${\rm t\bar{t} \to b\bar{b}\ell\nu_{\ell}\ell'\nu_{\ell'}}$), all-jets (${\rm t\bar{t} \to b\bar{b}qq'q''q'''}$) and $\ell$+jets (${\rm t\bar{t} \to b\bar{b}\ell\nu_{\ell}qq'}$) events. More information about the production and decay of ${\rm t\bar{t}}$ events can be found in Ref.~\cite{TTbarXS}.

\section{Direct measurements of the top-quark mass}

\subsection{Measurement of the top-quark mass in the dilepton channel}

The $m_{\rm t}$ measurement from CMS in the dilepton channel is performed with 2.3~fb$^{-1}$ of data~\cite{DileptonCMS}. Events in this channel are selected by requiring exactly two leptons (electrons (e) or muons ($\mu$)) with $p_{\rm T} > 20$~GeV and $|\eta| < 2.4$, at least 2 jets with $p_{\rm T} > 30$~GeV and $|\eta| < 2.4$, missing transverse energy $\not\!\!\!E_{\rm T} > 30$~GeV, and at least one b-tagged jet. In the ee and $\mu\mu$ channels events with $76 {\rm~GeV} < m_{\ell\ell} < 106$~GeV are rejected. The event-by-event top-quark mass $m_{\rm KINb}$ is reconstructed with the KINb algorithm~\cite{KINb}. For each of the possible jet-quark assignments the kinematic equations are solved multiple times per event, each time varying the reconstructed kinematics within their resolutions. The jet-quark assignment with the largest number of solutions is selected. Finally, $m_{\rm KINb}$ is extracted by taking the mean of a Gaussian fit to the distribution of the reconstructed top quark mass for all the different solutions of the kinematic equations, for the chosen jet-quark assignment.

The extraction of $m_{\rm t}$ is then performed by applying the template method. Templates which are sensitive to $m_{\rm t}$ are constructed for different top-quark mass hypotheses. The value of $m_{\rm t}$ is then extracted by doing a maximum likelihood fit of these templates to the distribution of $m_{\rm KINb}$ observed in data. By applying this technique, CMS measures $m_{\rm t} = 173.3 \pm 1.2 {\rm (stat)} ^{+2.5}_{-2.6} {\rm (syst)}$~GeV, where the systematic uncertainty is dominated by the global Jet Energy Scale (JES) uncertainty and the uncertainty on the flavour-dependent JES. This is the most precise measurement of $m_{\rm t}$ in the dilepton channel, with similar precision as the most recent result from D0~\cite{DileptonD0}.

\subsection{Measurement of the top-quark mass in the all-jets channel}

ATLAS performed the first measurement at the LHC of $m_{\rm t}$ in the all-jets channel, using 2.04~fb$^{-1}$ of data~\cite{AllHadrATLAS}. Events are selected by asking $\geq 5$ jets with $p_{\rm T} > 55$~GeV and $|\eta|<4.5$, and a 6$^{\rm th}$ jet with $p_{\rm T} > 30$~GeV and $|\eta|<4.5$. Two of these jets, with $p_{\rm T} > 55$~GeV and inside the inner detector acceptance ($|\eta|<2.5$), must be b-tagged. A cut on the $\not\!\!E_{\rm T}$ significance is also applied: $\not\!\!E_{\rm T} / \sqrt{H_{\rm T}} < 3$, where $H_{\rm T}$ is the scalar sum of the $p_{\rm T}$ of all the jets in the events. The ${\rm t \bar{t}}$ event topology is reconstructed using a 'mass $\chi^2$':
\begin{equation}
  \chi^2 = \frac{(m_{j_1,j_2} - m_{\rm W})^2}{\sigma^2_{\rm W}} + \frac{(m_{j_1,j_2,b_1} - m_{\rm t})^2}{\sigma^2_{\rm t}} + \frac{(m_{j_3,j_4} - m_{\rm W})^2}{\sigma^2_{\rm W}} + \frac{(m_{j_3,j_4,b_2} - m_{\rm t})^2}{\sigma^2_{\rm t}},
\end{equation}
where $\sigma_{\rm W} = 10.2$~GeV and $\sigma_{\rm t} = 17.4$~GeV are the respective mass resolutions from simulation. For every event, the jet-quark assignment with the lowest $\chi^2$ is taken, requiring $50 {\rm~GeV} < m_{j_1,j_2} < 110$~GeV and $50 {\rm~GeV} < m_{j_3,j_4} < 110$~GeV. This $\chi^2$ is minimized as a function of $m_{\rm W}$ and $m_{\rm t}$. Only events with $\chi^2_{\rm min} < 8$ are considered in the extraction of $m_{\rm t}$.

The template method is applied to the distribution of $m_{jjb}$ values from the selected jet-quark assignment, as observed in data. Each selected event contributes two values to this distribution. ATLAS measures $m_{\rm t} = 174.9 \pm 2.1 ({\rm stat}) \pm 3.8 ({\rm syst})$~GeV. The systematic uncertainty is dominated by the uncertainty on Initial and Final State Radiation (ISR/FSR), the uncertainty on the data-driven multijet background and the JES uncertainty.

\subsection{Measurement of the top-quark mass in the $\ell$+jets channel}

Both ATLAS and CMS have measured $m_{\rm t}$ in the $\ell$+jets channel~\cite{LeptonJetsATLAS,LeptonJetsCMS}. ATLAS selects events by asking exactly 1 isolated electron ($E_{\rm T} > 25$~GeV) or muon ($p_{\rm T} > 20$~GeV). The events need to have $\geq 4$ jets with $p_{\rm T} > 25$~GeV and $|\eta| < 2.5$, of which at least one is b-tagged. Multijet events are rejected by asking $\not\!\!\!E_{\rm T} > 35$~GeV and $m^{\rm T}_{\rm W} > 25$~GeV (e+jets), or $\not\!\!\!E_{\rm T} > 20$~GeV and $\not\!\!\!E_{\rm T} + m^{\rm T}_{\rm W} > 60$~GeV ($\mu$+jets). ATLAS is using two different approaches to measure $m_{\rm t}$, designed to reduce the JES uncertainty. Both analyses are based on the template method.

In the 1d-analysis, $R_{32} \equiv \frac{m_{\rm t}^{\rm reco}}{m_{\rm W}^{\rm reco}}$ is calculated for every event, where $m_{\rm t}^{\rm reco}$ and $m_{\rm W}^{\rm reco}$ are the reconstructed invariant masses of the hadronically decaying top quark and W boson, respectively. A kinematic fit is used to select a jet-quark assignment. The likelihood $L$ of the kinematic fit of the selected jet-quark assignment has to pass: $\ln{L} > -50$. The jets assigned to the $\rm t \to bqq'$ decay need to have $p_{\rm T} > 40$~GeV, and $m_{\rm W}^{\rm reco}$ has to fulfill $60 {\rm ~GeV} < m_{\rm W}^{\rm reco} < 100$~GeV.

The 2d-analysis, a combined measurement of $m_{\rm t}$ and a global Jet energy Scale Factor (JSF), is performed by a template fit to $m_{\rm t}^{\rm reco}$ and $m_{\rm W}^{\rm reco}$. In this analysis, the jet triplet assigned to the hadronic top decay is chosen as the one with maximum $p_{\rm T}$, considering only triplets with $50 {\rm ~GeV} < m_{\rm W}^{\rm reco} < 110$~GeV. Finally a kinematic fit is performed to the chosen jet triplet.

Both analyses are applied on 1.04~fb$^{-1}$ of data. The 2d-analysis has a slightly smaller uncertainty: $m_{\rm t} = 174.5 \pm 0.6 ({\rm stat}) \pm 2.3 ({\rm syst})$~GeV. The systematic uncertainty is dominated by the uncertainty on ISR/FSR and the uncertainty on the b-jet energy scale.

The CMS measurement of $m_{\rm t}$ uses 4.7~fb$^{-1}$ of data in the $\mu$+jets channel~\cite{LeptonJetsCMS}. Events are selected by asking exactly 1 isolated muon with $p_{\rm T} > 30$~GeV and $|\eta| < 2.1$, $\geq 4$ jets with $p_{\rm T} > 30$~GeV and $|\eta| < 2.4$, of which at least 2 are b-tagged. For every possible jet-quark assignment, a kinematic fit is applied with 3 mass constraints: an equal-mass constraint of $m_{\rm t}^{\rm lept}$ and $m_{\rm t}^{\rm hadr}$, and 2 $m_{\rm W}$ constraints. The kinematic fit returns the fitted top-quark mass $m_{{\rm t},i}^{\rm fit}$ and the fit probability $P_{\rm fit}^i$. Wrong jet-quark assignments are rejected by asking $P_{\rm fit}^i > 0.2$.

To extract $m_{\rm t}$ from the data, the Ideogram method is used in a combined measurement of $m_{\rm t}$ and the Jet Energy Scale (JES). In this method, a likelihood is calculated for every event:
\begin{equation}
  \mathcal{L} ({\rm event} \mid m_{\text{t}},\text{JES}) = \left( \sum_{i=1}^{n} P_{\rm fit}^i \cdot P \left( m_{{\rm t}, i}^{\rm fit}, m_{{\rm W},i}^{\rm reco} \mid m_{\rm t}, {\rm JES} \right) \right)^{\sum_{i=1}^n P_{\rm fit}^i}.
\end{equation}
The distributions $P \left( m_{{\rm t}, i}^{\rm fit}, m_{{\rm W},i}^{\rm reco} \mid m_{\rm t}, {\rm JES} \right)$ for all possible jet-quark assignments (correct assignments, wrong assignments and unmatched assignments) are taken from simulation. The individual event likelihoods are combined in a global likelihood, from which the measured $m_{\rm t}$ and JES values can be extracted. With this method, CMS measures $m_{\rm t} = 172.6 \pm 0.6 ({\rm stat}) \pm 1.2 ({\rm syst})$~GeV. The systematic uncertainty is dominated by the b-jet energy scale uncertainty and the uncertainty on the factorization scale. The systematic uncertainty does not include the uncertainty on color reconnection and on the underlying event.

\section{Measurement of the top-quark mass from the $\rm t \bar{t}$ cross section at $\sqrt{s} = 7$~TeV}

One of the problems of the direct measurements of $m_{\rm t}$ is that they use the mass definition from Monte Carlo generators, which is not related to $m_{\rm t}$  in a well-defined renormalization scheme ($m_{\rm t}^{\rm pole}$ or $m_{\rm t}^{\rm \overline{MS}}$) in a straightforward way. These masses can however be extracted from the measured cross section of top quark pair production $\sigma_{\rm t \bar{t}}$, since the theoretical dependency of $\sigma_{\rm t\bar{t}}$ on $m_{\rm t}^{\rm pole}$ or $m_{\rm t}^{\rm \overline{MS}}$ is known. ATLAS used $\sigma_{\rm t \bar{t}}$ measured from 35~pb$^{-1}$ of data in the $\ell$+jets channel, and obtains $m_{\rm t}^{\rm pole} = 166.4 ^{+7.8}_{-7.3}$~GeV for the pole mass~\cite{MassFromXSATLAS}. CMS used $\sigma_{\rm t\bar{t}}$ measured in the dilepton final state from 1.14~fb$^{-1}$ of data, and measures $m_{\rm t}^{\rm pole} = 170.3 ^{+7.3}_{-6.7}$~GeV and $m_{\rm t}^{\rm \overline{MS}} = 163.1 ^{+6.8}_{-6.1}$~GeV for the pole mass and the $\rm \overline{MS}$ mass, respectively~\cite{MassFromXSCMS}. Both the ATLAS and CMS results are in agreement with previous measurement performed by CDF and D0~\cite{TevTopMassTalk}.

\section{Measurement of the mass difference between top and antitop quarks}

One of the fundamental symmetries in the standard model, the invariance under CPT transformations, can be tested by measuring the difference in mass between a particle and the corresponding antiparticle. Since the top quark is the only quark that decays before hadronization can take place, this difference can be measured directly. CMS performed a measurement using 1.09~fb$^{-1}$ of data in the $\mu$+jets channel~\cite{TopMassDiff}. The events were splitted in two distinct samples according to the charge of the lepton. In each of these two samples, the mass of the hadronically decaying top quarks was measured and finally both masses were subtracted from eachother. This resulted in $\Delta m_{\rm t} = -1.20 \pm 1.21 ({\rm stat}) \pm 0.47 ({\rm syst})$~GeV. The smallness of the systematic uncertainty, when compared to $m_{\rm t}$ measurements, can be explained by the cancellation of most systematics by taking the difference.

\section{Conclusion and outlook}

\begin{figure}[htb]
  \centering
  \includegraphics[width=0.6 \textwidth]{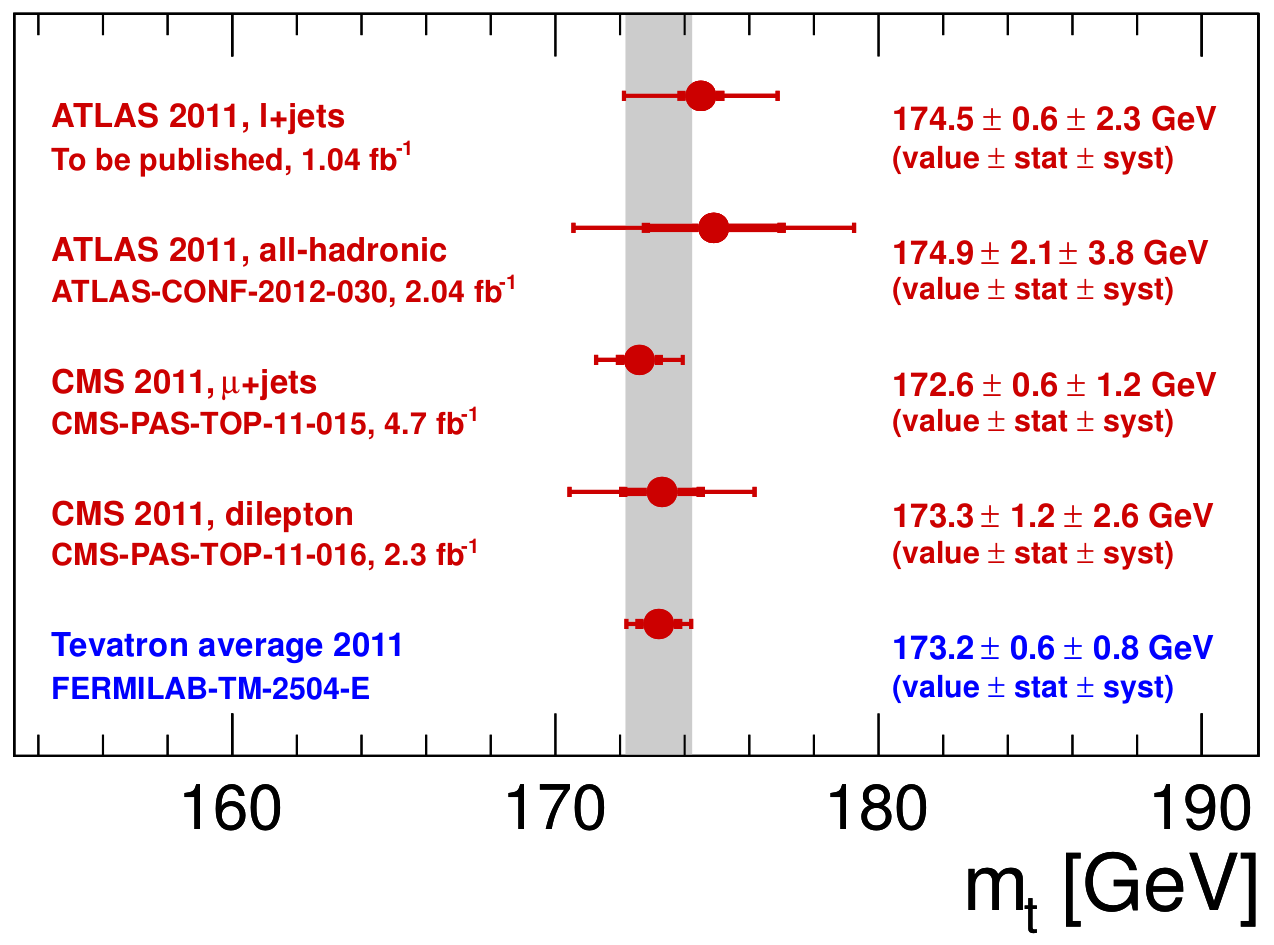}
  \caption{Overview of the $m_{\rm t}$ measurements from ATLAS and CMS, including the latest CDF and D0 combination. \label{fig:OverviewMtop}}
\end{figure}

Currently the most precise measurements of $m_{\rm t}$ are performed by the Tevatron experiments, but the ATLAS and CMS results are getting more and more precise. An overview of all these results can be found in Fig.~\ref{fig:OverviewMtop}.
The next step is to combine all the CMS and ATLAS measurements, on which work has already started, and finally to combine these also with the CDF and D0 results. CMS already performed a combination of its measurements, resulting in $m_{\rm t} = 172.6 \pm 0.4 (\text{stat}) \pm 1.2(\text{syst})$~GeV.

With the huge amount of data which will be recorded by ATLAS and CMS, the precision on $m_{\rm t}$ is expected to increase. The uncertainties are currently dominated by the jet energy scale systematic uncertainty, which can be reduced once more data is available to be analyzed.


\section*{References}
\begin{thebibliography}{99}
\bibitem{TevatronCombi}CDF and D0 Collaboration, ``Combination of CDF and DO results on the mass of the top quark using up to 5.8~fb$^{-1}$ of data'', (2011). \texttt{arXiv:1107.5255}.

\bibitem{TevTopMassTalk}O. Brandt, ``Measurements of the top quark mass at the Tevatron'', these proceedings. \texttt{arXiv:1204.0919}.

\bibitem{TTbarXS}I. Aracena, ``Top Pair Production at $\rm E_{cm} = 7$~TeV'', these proceedings.

\bibitem{DileptonCMS}CMS Collaboration, ``Measurement of the top-quark mass in the dilepton channel in pp collisions at $\sqrt{s} = 7$ TeV'', {\it CMS Physics Analysis Summary} {\bf CMS-PAS-TOP-11-016} (2012).

\bibitem{KINb}CMS Collaboration, ``Measurement of the ${\rm t \bar{t}}$ production and the top-quark mass in the dilepton channel in pp collisions at $\sqrt{s} = 7$~TeV'', {\it JHEP} {\bf 07} (2011) 049. \texttt{arXiv:1105.5661}.

\bibitem{DileptonD0}D0 Collaboration, ``Measurement of the top quark mass in ${\rm p \bar{p}}$ collisions using events with two leptons'', submitted to {\it Phys. Rev. Lett.} \texttt{arXiv:1201.5172}.

\bibitem{AllHadrATLAS}ATLAS Collaboration, ``Determination of the Top Quark Mass with a Template Method in the All-Hadronic Decay channel using 2.04~fb$^{-1}$ of ATLAS data'', {\it ATLAS Note} {\bf ATLAS-CONF-2012-030} (2012).

\bibitem{LeptonJetsATLAS}ATLAS Collaboration, ``Measurement of the top quark mass with the template method in the top antitop $\to$ lepton + jets channel using ATLAS data'', submitted to {\it Eur. Phys. J.} {\bf C}, \texttt{arXiv:1203.5755}.

\bibitem{LeptonJetsCMS}CMS Collaboration, ``Measurement of the top quark mass in the muon+jets channel'', {\it CMS Physics Analysis Summary} {\bf CMS-PAS-TOP-11-015} (2012).

\bibitem{MassFromXSATLAS}ATLAS Collaboration, ``Determination of the Top-Quark Mass from the $\rm t\bar{t}$ Cross Section Measurement in pp Collisions at $\sqrt{s} = 7$~TeV with the ATLAS detector'', {\it ATLAS Note} {\bf ATLAS-CONF-2011-054} (2011).

\bibitem{MassFromXSCMS}CMS Collaboration, ``Determination of the Top Quark Mass from the $\rm t \bar{t}$ Cross Section at $\sqrt{s} = 7$~TeV'', {\it CMS Physics Analysis Summary} {\bf CMS-PAS-TOP-11-008} (2011).

\bibitem{TopMassDiff}CMS Collaboration, ``Measurement of the mass difference between top and antitop quarks'', {\it CMS Physics Analysis Summary} {\bf CMS-PAS-TOP-11-019} (2011).

\end{thebibliography}
\end{document}